\documentclass[aip,jcp,amsmath,amssymb,twocolumn,reprint]{revtex4-1}
\usepackage{hyperref}
\usepackage{ulem}

\usepackage{amsmath, amssymb, amsfonts}
\usepackage{graphicx}
\usepackage{float}
\usepackage{siunitx}

\usepackage{mhchem}
\newcommand{\be}{\begin{equation}}
\newcommand{\ee}{\end{equation}}
\newcommand{\bea}{\begin{eqnarray}}
\newcommand{\eea}{\end{eqnarray}}
\newcommand{\Eq}{Eq.~\ref}
\newcommand{\Fig}{Fig.~\ref}

\renewcommand\Re{\operatorname{Re}}

\begin{document}
%
\title{Comment on ``Benchmarking Compressed Sensing, Super-Resolution, and Filter Diagonalization''}
%
%
\author{Vladimir A. Mandelshtam}

\affiliation{Chemistry Department,
University of California at Irvine, CA 92697, USA}

\date{\today}

\begin{abstract}
In a recent paper [{\it Int. J. Quant. Chem.} (2016) DOI:
  10.1002/qua.25144] Markovich, Blau, Sanders, and Aspuru-Guzik presented a numerical evaluation and
  comparison of three methods,
  Compressed Sensing (CS), Super-Resolution (SR), and Filter Diagonalization
  (FDM),  on their ability of ``recovering information'' from time
  signals, concluding that CS and RS outperform FDM. We argue that
  this comparison is invalid for the following reasons. FDM is a well
  established method designed for solving the harmonic inversion problem or,
  similarly, for the problem of spectral estimation, and as such
  should be applied only to problems of this kind. The authors
  incorrectly assume that the problem of data fitting is equivalent to the
  spectral estimation problem, regardless of what parametric form is
  used, and, consequently, in all five numerical examples FDM is applied to
  the wrong problem. Moreover, the authors' implementation of FDM turned out
  to be incorrect, leading to extremely bad results, caused by
  numerical instabilities. As we demonstrate here, if implemented correctly, FDM could
  still be used for fitting the data, at least for the time signals
  composed of damped sinusoids, resulting in superior performance. In
  addition, we show that the published article is full of
  inaccuracies, mistakes and incorrect
statements.

\end{abstract}

\maketitle

Given a (generally complex-valued) time signal sampled on an
equidistant time grid $f_n:=f(n\tau)$,
the Filter Diagonalization Method (FDM) \cite{wall1995,mandelshtam1997,hu1998,mandelshtam2001} is designed to solve
the Harmonic Inversion Problem (HIP),

\be\label{eq:HIP}
\sum_{j=1}^K \lambda_j u_j^n=f_n,\ \ \ (n=0,...,N-1),
\ee
 for the unknown (generally complex) amplitudes $\lambda_j$ and poles $u_j\equiv
 e^{-i\tau(\omega_j-i\gamma_j)}$, where $\omega_j$ are frequencies
 and $\gamma_j$ are the decay parameters. (It is usually assumed that
 all $\gamma_j\ge 0$, so that the time signal does not increase
 exponentially with $n$.)

In FDM the seemingly nonlinear fitting
problem  \eqref{eq:HIP} is solved by mapping it to a
generalized eigenvalue problem with data matrices defined by the
sequence $\{f_n\}$. If the condition ({\bf the information uncertainty principle})
\be\label{eq:cond}
N\ge 2K
\ee
 is satisfied with
exact arithmetics, then FDM provides the {\bf exact} solution of HIP \eqref{eq:HIP}. To
a certain extent, this property of FDM is shared with other linear
algebraic methods (Linear Regression, Linear Prediction, Matrix
Pencil, etc.), with the differences buried in the
details. From a numerical/algorithmic perspective, FDM is
perhaps the best such method that combines numerical efficiency with
robustness and accuracy, and as such demonstrates nearly optimal
performance. One important and distinct
feature of FDM is the use of a Fourier basis,
which allows for the reduction of a large and often ill-conditioned
generalized eigenvalue problem to a set of small and well-conditioned
 problems, effectively reducing the global parametric
fit \eqref{eq:HIP} to a local spectral analysis. Consequently, in practice condition \eqref{eq:cond}
holds in a weak sense, especially for large data sets, when only local
spectral analysis can be performed, in which case it reads as
\be\label{eq:cond1}
N\ge {4\pi\rho }/{\tau},
\ee
where $\rho$ is the local density of states (i.e., the density of 
frequencies $\omega_j$).

Consider the infinite-time discrete Fourier transform
(FT) of the time signal $f_n$ satisfying \Eq{eq:HIP}:
\be\label{eq:FT}
I(\omega):=\tau \sum_{n=0}^{\infty} f_n z^{-n}=\tau\sum_j \frac
{\lambda_j}{1-u_j/z};\ \ \ (z:=e^{-i\tau\omega})
\ee
This function of frequency has peaks centered at 
$\omega=\omega_j$ with amplitudes given by $\lambda_j$, and as such
provides the
information related to the ``line list''  $\{\lambda_j,u_j\}$.
(Note also that for a ``phased'' signal, e.g., assuming all the amplitudes $\lambda_j$
being real, the {\bf absorption mode} spectrum, $\Re I(\omega)$, has line shapes
superior to those of the absolute value spectrum $|I(\omega)|$.)
Consequently, the parametric fit problem \eqref{eq:HIP} is related to
the spectral estimation problem, i.e., given a finite sequence
$\{f_n\}$ ($n=0,...,N-1$) estimate its infinite-time FT $I(\omega)$.

The conventional way to estimate $I(\omega)$ is to use the finite-time FT
\be\label{eq:FT}
I^{\rm (FT)}(\omega):=\tau \sum_{n=0}^{N-1} f_n g_n z^{-n} 
\ee
where $g_n$ is a suitable apodization function. As is well
known, the main drawback of \Eq{eq:FT} is its slow convergence with
respect to the data size $N$, often referred to as the {\bf FT uncertainty
principle} for the spectral resolution $\delta\omega$ as a function of signal size:
\be\label{eq:uncert}
\delta\omega \sim 2\pi/ {N\tau}.
\ee

According to \Eq{eq:FT}, the line list determined from solving HIP
\eqref{eq:HIP} can be used to estimate $I(\omega)$ directly.
Consequently, the FDM spectral estimate may result in high (or even
infinite) resolution (depending on how well 
condition \eqref{eq:cond1} is satisfied), while the resolution of the
finite FT spectrum is still limited by the uncertainty
principle \eqref{eq:uncert}. 

Importantly, FDM never attempts to solve the spectral estimation
problem by  first {\bf extrapolating} the time signal $f_n$
to longer times ($n>N$) followed by its Fourier transformation. In fact, the extrapolation problem is
generally much harder than the
problem of spectral estimation.

All of the above has been explained and demonstrated in a number of
publications (see, e.g.,
refs. \cite{wall1995,mandelshtam1997,hu1998,mandelshtam2001,martini2014}
to mention just a few). 
Of course things become much more complicated and less
straightforward when the data in question has imperfections, i.e.,
when the benefits from the solution of HIP \eqref{eq:HIP} become
unclear. In this regard we note ref. \cite{martini2014}, in
which FDM performance was critically assessed by applying it to noisy
data, both synthesized and experimental. 

Quite surprisingly, the authors of the recent publication
\cite{markovich2016}, which from now on we refer to as {\it the
  Paper}, managed to ignore all of the knowledge accumulated
since 1997, when the FDM algorithm in its present form was first
introduced \cite{mandelshtam1997}. It reports a study
comparing three methods: Super-Resolution
(SR), Compressed Sensing (CS), and FDM.
{\it The Paper} is full of various inaccuracies, mistakes, and incorrect
statements and assumptions, starting already in  
the abstract: {\it ``Signal processing
techniques have been developed that use different strategies to bypass
the {\bf Nyquist sampling theorem} in order to recover more information than
a traditional discrete Fourier transform.''} In light of what we have
discussed above, in the abstract and later in {\it the Paper} the
authors probably mean {\it the FT uncertainty principle} \eqref{eq:uncert},
which is indeed the main
motivation for developing various super resolution techniques, such as
FDM, while the mentioned ``sampling theorem'' simply refers to the
relationship between the time interval $\tau$ for a uniformly
sampled data, $f_n=f(n\tau)$, and the range of recoverable
frequencies (the Nyquist range):
\be \label{eq:nyquist}
[\omega_{\rm min} ; \omega_{\rm max}] =[-{\pi}/{\tau};{\pi}/{\tau}]
\ee
The corresponding phenomenon associated with this theorem is often
referred to as ``aliasing'' or ``folding'', and the only way to circumvent it with
uniformly sampled data is to reduce the sampling time $\tau$. 

A comparison of FDM with any other signal processing technique would be
meaningful only within the framework discussed above, i.e., by comparing
the performance of the methods as either parameter estimators or 
spectral estimators. It may appear that the authors' original
intention was to carry out such a comparison, but instead they decided
to solve
the problem of {\bf data-fitting} without addressing either HIP
\eqref{eq:HIP}  or the spectral estimation problem. In one sentence,
{\it the Paper} demonstrates that a violin (FDM) is not good for
chopping vegetables. In order to justify their way to
assess the performance of the three methods the authors refer to
``Parseval's theorem'' (page 4), claiming that a good fit in the
time domain implies a good fit in the frequency domain. This statement
is incorrect. In fact, even a perfect fit of the data within a finite time
interval does not solve the FT spectral estimation problem, unless a very
specific parametrization, directly related to the latter is used, and
FDM is designed to accomplish this goal, while neither of the five
numerical examples in {\it the Paper} test the methods in question for
their ability in spectral estimation.

The first numerical example is the most extreme in terms of its
irrelevance to the problem of spectral estimation. It considers a Gaussian
time signal (Eq. 24 in  {\it the Paper}), i.e. a time signal with no
spectral features. Both CS and SR then attempt to
expand $f(t)$ as a linear combination of 10,000  Gaussian basis functions with one of the
basis functions being {\bf identical} to $f(t)$. Not surprisingly,
the resulting fit of $f(t)$ as a ``linear combination of Gaussians''
is perfect! The question is: what goal has been accomplished by
fitting a single Gaussian with  itself? 

Interestingly, on several occasions, including the corresponding statement in the abstract, the
authors try to compliment FDM on {\it ``providing the best results for
Lorentzian signals''}, however, their numerical examples prove the
opposite: the error estimates for FDM, in all five numerical examples, are so big compared to that
for CS and SR, that a logarithmic scale is required, even for the time signals
exactly satisfying the form of HIP \eqref{eq:HIP}, for which,
according to all our experience, FDM should provide numerically exact results. (Plotting the results obtained by 
CS and SR in the same plot with the authors' version of the FDM results achieved only
one goal: they made both
SR and CS look very good, even
when their errors were not small.)

On page 3 the authors describe Filter Diagonalization. Several
mathematical expressions contain unrecognized symbols. Before Eq. 11, they
incorrectly call the evolution operator {\it unitary}. Most of the
citations in this section are incorrect. For example, the {\it ``2D filter diagonalization''} was not
introduced in [R. Chen, H. Guo, J. Chem. Phys. 1999, 111, 464], and
{``multi-resolution filter diagonalization''} claimed to be introduced in
[K. Aizikov, P. B. O’Conn\underline{e}r, J. Am. Soc. Mass Spectrom. 2006, 17, 836]
does not exist. The {\it ``adaptive frequency grid''} was proposed
in ref. \cite{mandelshtam1997}, but was not recommended in our later
publications. In fact, a very simple uniform grid 
with spacings given by
\be
\Delta \nu={4\pi}/ {(N\tau)}
\ee
defines the Fourier basis
with optimal performance (see
e.g. refs. \cite{hu1998,mandelshtam2001,martini2014}). Apparently, in
their implementation of FDM
these authors used a frequency grid in the range of 0 to 20 kHz
(or even 0 to 200kHz), while according to the Nyquist criterion
\eqref{eq:nyquist} the maximum meaningful
frequency would be $\omega_{\rm max}=4095\pi$ rad/s ($\tau=1/4095$ s), i.e. they used the
frequency range an order of
magnitude larger than the correct range. (For the undersampled data
(see below) the
discrepancy is even greater.) Moreover, they then attempted
to use 
an {\bf adaptive frequency grid}.  We believe that this poor choice of
the frequency grid defining the FT basis
was the main reason for the numerical instabilities (mentioned on page
8) encountered by these authors. In light of this instability,
it is unclear how they {\it ``ensured the robustness of
their results''} (as they mentioned on page 4). 

There are many other inconsistencies and/or mistakes and/or
inaccuracies in {\it the Paper}. In the remaining part of this Comment
we mention the ones which are most striking. 

\begin{figure}
	\includegraphics[width=1.0\columnwidth]{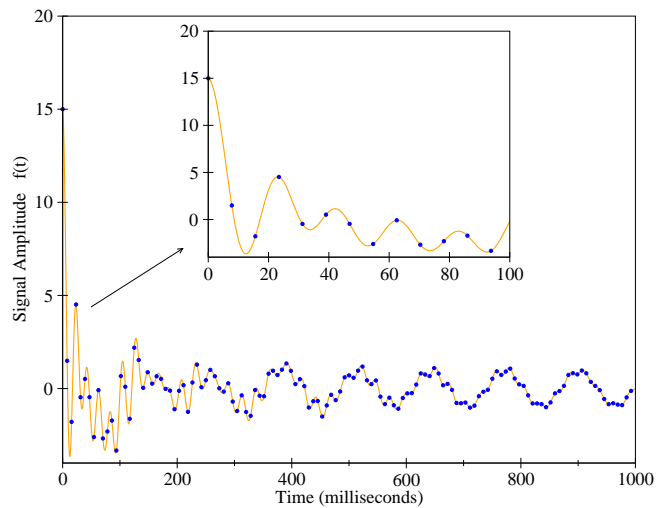}
	\caption{ \label{fig:signal}
	The time signal $f(t)$ from Fig 5a from {\it the Paper},
        defined here in the caption of Table \ref{tab:par}. The 128 blue dots
        indicate the data points used to process $f_n=f(n\tau)$ by FDM
        with $\tau=1/127$s.
	}
\end{figure}

Eq. 28 in {\it the Paper} with 20 random damped cosines 
\be
f(t)=\sum_{n=1}^{20}e^{-\gamma_n t} \cos \omega_n t\ \ \ \ \mbox{(Eq. 28 in {\it the Paper}) }
\ee
is supposed to
define the time signal shown in Fig. 5a of {\it the Paper}.
The description of this time signal says that the frequencies $\omega_n$
are drawn randomly from the range  between 0 to 50$\pi$ Hz.
Here in \Fig{fig:signal}  we reproduce this time signal. 
By examining the plot of $f(t)$ we find that $f(0)=15$, and not 20! That is, the data shown in  {\it
  the Paper} is not consistent with its description. The power
spectrum in Fig. 5b of {\it the Paper} does have about 20 peaks, but
the spectral range is $[0;2048]$Hz rather than $[0;50\pi]$Hz. Table
\ref{tab:par} lists the parameters $\{\omega_n,\gamma_n\}$
($n=1,...,15$) of $f(t)$ shown in the figure, and the maximum
frequency, $\omega_{15}=304.95$ rad/s, is outside of the range ($[0;50\pi]$Hz)
indicated in {\it the Paper}. Regardless of all these inconsistencies,
the time signal from \Fig{fig:signal} with the frequencies $\omega_n$
and decays $\gamma_n$ from the table represents an easy toy problem for our
old FDM code written in Fortran 77 in 1997 \cite{mandelshtam1997}. (Note that
the sum of 15 damped cosines corresponds to a sum of 30 damped complex
exponentials.)

\begin{figure}
	\includegraphics[width=1.0\columnwidth]{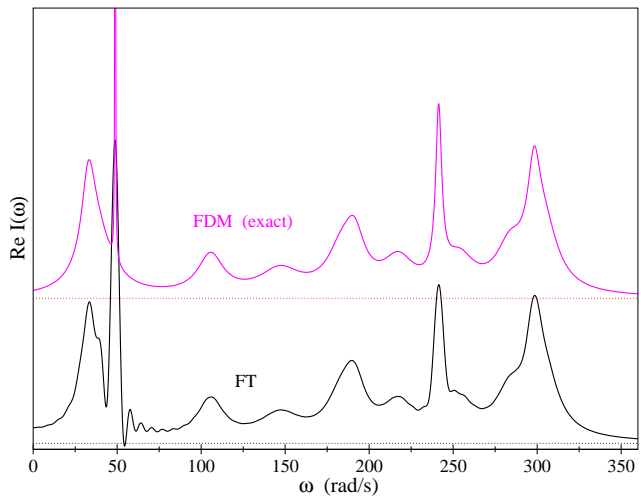}
	\caption{ \label{fig:spectra}
	The FT and FDM spectral estimates of $\Re I(\omega)$ of the
        time signal $f(t)$ shown in Fig 5a in {\it the Paper} and here
        in \Fig{fig:signal}.
The FT is evaluated using the entire data set consisting of 4096
points, and the FDM spectrum (which coincides with the exact spectrum)
 is computed using the 128 points indicated in \Fig{fig:signal} by
 blue dots. The FDM spectrum using only 64 first data points (out of
 128) would also be indistinguishable from the exact result if shown in this plot.
	}
\end{figure}

Here is how the authors describe their numerical protocol (page 4):\\
{\it ``For each sparse signal processing method, we varied how many of
  the 4096 time points we sampled (in increments of 64) and
  investigated the dependence of the recovery error on the extent of
  undersampling.''} \\
Then later on the same page:
{\it ``Because super-resolution requires a grid of equally spaced
  sample points, we began our analysis by examining the full signal
  and computing the errors. We then repeated our analysis for the
  signal by successively undersampling in powers of two, taking care
  to ensure that our sample points were always equally spaced.''}\\
These two contradictory statements do not explain how the data was
actually ``undersampled''.
Thus, we make a reasonable assumption and apply FDM to the signal from
\Fig{fig:signal} using only 128
points, i.e., taking every 32-nd point (as
shown by blue dots). This corresponds to the time step $\tau=1/127$s
and the Nyquist range $[\omega_{\rm min} ; \omega_{\rm max}]=[-127\pi;
127\pi]$, i.e., just enough to correctly reconstruct the spectrum.
 A solution of the small generalized
eigenvalue problem then gave exactly
30 complex frequencies and amplitudes that agreed with those in the
table to many significant figures. That is, we confirmed once again, that the solution of  HIP by
FDM in the case of perfect Lorentzian signal is numerically exact. As
expected, no numerical instability has been encountered. We then used the computed
line list to construct the spectrum $\Re I(\omega)$ as
shown in \Fig{fig:spectra}. Moreover, nearly the same spectral estimate
(not shown) is
obtained by using only the first $N=64$ data points (out of 128) of the same
subset,
although in the latter case some of the entries in the line list were
not extremely accurate. Note that the FT spectrum computed using all the 4096 
signal points shows some artifacts (Gibbs oscillations) around the
narrow feature at $\omega=48.82$ rad/s.

\begin{table*}
\caption{The frequencies $\omega_n$ and decays $\gamma_n$ (in rad/s) of the time
  signal, $f(t)=\sum_{n=1}^{15}e^{-\gamma_n t} \cos \omega_n t$, shown in \Fig{fig:signal}.
}
\bigskip

\label{tab:par}
\begin{tabular}{|c|c|c|c|c|c|c|c|c|c|c|c|c|c|c|c|}
\hline
$\omega_n$ & 32.85 & 36.51 & 38.29 & 48.82 & 105.63 &146.96 & 183.24 & 191.10
&217.34 & 241.39 & 253.59 & 283.60 & 298.20 &299.85 &304.95  \\
\hline
$\gamma_n$ & 4.61   & 9.44   & 12.95 & 0.09   & 9.24     & 14.88  & 10.19   & 7.46
& 11.21 & 2.18 & 11.03 & 9.64 & 4.42 & 12.25 & 9.20 \\
\hline

\end{tabular}
\end{table*}

In conclusion, we have demonstrated that {\it the Paper} cannot be
considered a valid ``benchmarking'' of any of the methods in
question. 
Based on the critical mass of direct and indirect evidence described above, we
conclude that the numerical
implementation of FDM by these authors was done poorly,
ignoring most of the ``know how'' accumulated in a number of publications
since 1997.
While FDM has already established itself as an efficient spectral estimator, both CS
and RS still require a proper investigation. Note that two earlier papers by
related groups of authors
presented CS \cite{andrade2012} and RS \cite{markovich2013}  as
methods for FT
spectral reconstruction of time signals using data from molecular
dynamics simulations. That is, unlike {\it the Paper} these two publications were
concerned with a correct problem. In ref. \cite{andrade2012} the authors
concluded that CS beats FT by a factor of {\bf five}  and in
ref. \cite{markovich2013}  they gave a slightly less optimistic factor
({\bf four}) by which RS beats FT.  
Both conclusions were made
based on a single application of the method in question to
irreproducible short-time data, and most importantly, with the
converged spectrum unavailable. (In both cases, all of the spectra involved
in the comparisons were not converged.) Therefore, the
usefulness of either CS or SR for spectral reconstruction remains to
be demonstrated.

\bigskip
\noindent
{\bf Acknowledgments.}\hskip 0.2in We thank Thomas Markovich for
sharing his data with us, and all the authors of {\it the
  Paper} for acknowledging ``useful conversations with Vladimir Mandelshtam''.
The NSF support, Grant No. CHE-1566334, is acknowledged.

\end{document}